\documentclass[showpacs,aps,prd,reprint,superscriptaddress,nofootinbib,longbibliography]{revtex4-1}
\usepackage[colorlinks=true, pdfstartview=FitV,
bookmarks=true, bookmarksnumbered=true, breaklinks]{hyperref}
\usepackage[dvipdfmx]{graphicx}
\usepackage{amsmath,amssymb,bm,color,longtable,mathrsfs,slashed,comment,tikz}
\definecolor{darkpastelgreen}{rgb}{0.01, 0.75, 0.24}
\definecolor{electricindigo}{rgb}{0.44, 0.0, 1.0}
\definecolor{palatinateblue}{rgb}{0.15, 0.23, 0.89}
\definecolor{carminered}{rgb}{1.0, 0.0, 0.22}
\hypersetup{linkcolor=palatinateblue, citecolor=electricindigo, urlcolor=carminered}

\definecolor{lime}{HTML}{A6CE39}
\DeclareRobustCommand{\orcidicon}{
	\hspace{-3mm}
	\begin{tikzpicture}
	\draw[lime, fill=lime] (0,0) 
	circle [radius=0.16] 
	node[white] {{\fontfamily{qag}\selectfont \tiny ID}};
	\draw[white, fill=white] (-0.0625,0.095) 
	circle [radius=0.007];
	\end{tikzpicture}
	\hspace{-3mm}
}

\foreach \x in {A, ..., Z}{\expandafter\xdef\csname orcid\x\endcsname{\noexpand\href{https://orcid.org/\csname orcidauthor\x\endcsname}
			{\noexpand\orcidicon}}
}

\newcommand{\Slash}[1]{{\ooalign{\hfil/\hfil\crcr$#1$}}}

\begin{document}
\begin{flushright}
\flushright{DESY~21-110,~KEK-TH-2338}
\end{flushright}

\title{Kondo effect with Wilson fermions}

\author{Tsutomu~Ishikawa\orcidA{}}
\email[]{tsuto@post.kek.jp}
\affiliation{The Graduate University for Advanced Studies (SOKENDAI), Tsukuba 305-0801, Japan}
\affiliation{KEK Theory Center, Institute of Particle and Nuclear
Studies, High Energy Accelerator Research Organization (KEK), Tsukuba 305-0801, Japan}
\affiliation{RIKEN Center for Computational Science, Kobe 650-0047, Japan}

\author{Katsumasa~Nakayama\orcidB{}}
\email[]{katsumasa.nakayama@desy.de}
\affiliation{NIC, DESY Zeuthen, Platanenallee 6, 15738 Zeuthen, Germany}

\author{Kei~Suzuki\orcidC{}}
\email[]{k.suzuki.2010@th.phys.titech.ac.jp}
\affiliation{Advanced Science Research Center, Japan Atomic Energy Agency (JAEA), Tokai 319-1195, Japan}

\date{\today}

\begin{abstract}
We investigate the Kondo effect with Wilson fermions.
This is based on a mean-field approach for the chiral Gross-Neveu model including four-point interactions between a light Wilson fermion and a heavy fermion.
For massless Wilson fermions, we demonstrate the appearance of the Kondo effect.
We point out that there is a coexistence phase with both the light-fermion scalar condensate and Kondo condensate, and the critical chemical potentials of the scalar condensate are shifted by the Kondo effect.
For negative-mass Wilson fermions, we find that the Kondo effect is favored near the parameter region realizing the Aoki phase.
Our findings will be useful for understanding the roles of heavy impurities in Dirac semimetals, topological insulators, and lattice simulations.
\end{abstract}

\maketitle

\section{Introduction} \label{Sec:1}
The Kondo effect has a long history in solid-state physics~\cite{Kondo:1964,Hewson,Yosida,Yamada,coleman_2015}.
It was observed as an enhancement of electric resistance of a metal, and it is induced by a strong correlation between nonrelativistic itinerant electrons and localized spin impurities.
Kondo effects can be also realized for {\it relativistic} fermions such as Dirac/Weyl/Majorana fermions.
Such relativistic Kondo effects can occur in relativistic-fermion systems including impurities, such as graphene (see Ref.~\cite{Fritz_2013} for a review), Dirac/Weyl semimetals~\cite{Principi:2015,Yanagisawa:2015conf,Yanagisawa:2015,Mitchell:2015,Sun:2015,Feng:2016,Kanazawa:2016ihl,Lai:2018,Ok:2017,PhysRevB.97.045148,PhysRevB.98.075110,Dzsaber:2018,PhysRevB.99.115109,KIM2019236,Grefe:2019,Grefe:2020,Pedrosa:2021}, dense nuclear matter~\cite{Yasui:2013xr,Yasui:2016ngy,Yasui:2016hlz,Yasui:2019ogk}, and dense quark matter~\cite{Yasui:2013xr,Hattori:2015hka,Ozaki:2015sya,Yasui:2016svc,Yasui:2016yet,Kanazawa:2016ihl,Kimura:2016zyv,Yasui:2017izi,Suzuki:2017gde,Yasui:2017bey,Kimura:2018vxj,Macias:2019vbl,Hattori:2019zig,Suenaga:2019car,Suenaga:2019jqu,Kanazawa:2020xje,Araki:2020fox,Araki:2020rok,Suenaga:2020oeu}.
Among them, the ``QCD Kondo effect"~\cite{Yasui:2013xr,Hattori:2015hka} is induced by the color exchange interaction between a light quark and an impurity quark, which is based on quantum chromodynamics (QCD).
To determine the parameter region (or phase diagram) realizing the QCD Kondo effect is one of the challenging problems in QCD.

In this paper, we focus on the Kondo effect for the Wilson fermion.
The Wilson fermion is one of the formulations realizing Dirac-like lattice fermions, which was first proposed in the viewpoint of construction of lattice gauge field theories~\cite{Wilson:1975,Wilson:1977}.
It has been very useful to implement quark degrees of freedom in lattice QCD simulations, and also approximate Wilson fermions can be realized in Dirac semimetals.

In particular, the negative-mass region of the Wilson fermion is physically interesting because a part of this region corresponds to the bulk mode of topological insulators.
In addition to the negative mass, an interaction between fermions, such as four-point or gauge interaction, can induce a new phase with spontaneous parity symmetry breaking for $N_f=1$ ($N_f$ is the number of flavors) or parity-flavor symmetry breaking for $N_f=2$, which is the so-called Aoki phase~\cite{Aoki:1983qi}.
The Aoki phase for the Wilson fermion was discussed by mean-field theories~\cite{Aoki:1983qi,Aoki:1985jj,Aoki:1987us,Aoki:1993vs,Horvath:1995da,Izubuchi:1998hy,Bermudez:2018eyh,Ziegler:2020zkq}.\footnote{See Refs.~\cite{Azcoiti:2008dn,Sharpe:2008ke,Azcoiti:2012ns,Azcoiti:2013oba} for arguments about an additional flavor-singlet condensate for $N_f=2$.}\footnote{The Aoki phase can appear in other lattice-fermionic systems such as the domain-wall fermion~\cite{Vranas:1999nx,Izubuchi:1999fk,Aoki:2000pc}, a naive or staggered fermion with a taste-splitting mass term~\cite{Creutz:2011cd}, staggered-Wilson fermions~\cite{Misumi:2012sp}, and minimal doubling fermions~\cite{Misumi:2012uu,Kamata:2013wka}.}
For QCD in continuum space, the Aoki phase is regarded as an artifact due to the discretization of the spacetime, but in solid-state physics, similar phase structures were pointed out by an interacting Su-Schrieffer-Heeger model~\cite{Kuno:2018pcp}, an interacting Kane-Mele model~\cite{Araki:2013dsa}, and a Fu-Kane-Mele-Hubbard model~\cite{Sekine:2014xva}.
Such parity-broken materials are also closely related to axion insulators (see Ref.~\cite{Sekine:2021} for a review).
In this work, we investigate the interplay between the Aoki phase and the Kondo effect.
Our studies will be useful for elucidating impurity effects in strongly correlated lattice fermion systems.

This paper is organized as follows.
In Sec.~\ref{Sec:2}, we construct our model.
In Sec.~\ref{Sec:3}, we show our numerical results and discuss properties of the Kondo effect with the Wilson fermion.
Section~\ref{Sec:4} is devoted to our conclusion and outlook.

\section{Formulation} \label{Sec:2}
The Kondo effect for high-momentum particles can be described as a perturbative scattering problem between a light fermion and a heavy impurity.
On the other hand, for the low-momentum region, the perturbative expansion does not converge, so that a nonperturbative approach is needed.
In order to investigate the nonperturbative Kondo effect and its competition with other nonperturbative effects, we employ a mean-field approach.
Mean-field approaches have been successfully applied to the conventional Kondo effect~\cite{Read:1983,Coleman:1983}, and similar approaches should be also used for relativistic fermions (for a model with Nambu--Jona-Lasinio (NJL)-type four-point interactions, see Refs.~\cite{Yasui:2016svc,Yasui:2017izi}).

For the light-fermion sectors, we use the ``chiral Gross-Neveu ($\chi$GN) model" in the $1+1$ dimensions~\cite{Gross:1974jv} (namely, the NJL$_2$ model), which includes not only the scalar-type four-point interaction but also the pseudoscalar-type one.
This model is used as a toy model for QCD.
After replacing the (Dirac-type) continuous fermion by the Wilson fermion, we call this model the ``Wilson-chiral-Gross-Neveu (W$\chi$GN) model."
This model was first studied in Ref.~\cite{Eguchi:1983gq}, and, for early studies about the Aoki phase, see Refs.~\cite{Aoki:1983qi,Aoki:1985jj} at zero chemical potential and Ref.~\cite{Izubuchi:1998hy} at nonzero chemical potential.

For the sectors including heavy-fermion fields, we introduce a heavy-fermion field based on the heavy-quark effective theory (HQET)~\cite{Eichten:1989zv,Georgi:1990um,Neubert:1993mb,Manohar:2000dt}.
Although this field is regarded as a heavy-mass limit of the original massive Dirac field, it should be valid as long as its mass scale is sufficiently larger than its other typical scales.
Furthermore, we use a four-point interaction between light and heavy fermions.
Even if such a heavy-light four-point interaction may be regarded as an approximate form of the underlying interaction, it can be applied to nonperturbative physics such as strongly coupled heavy-light bound states (namely, mesons)~\cite{Ebert:1994tv,Ebert:1996vx,Mota:2006ex,Guo:2012tm} and the Kondo effect~\cite{Yasui:2016svc,Yasui:2017izi}.

By combining the light- and heavy-fermion sectors, we can construct a ``Wilson-chiral-Gross-Neveu-Kondo (W$\chi$GNK) model."\footnote{Precisely speaking, this model is analogous to the Coqblin-Schrieffer model~\cite{Coqblin:1969} rather than the Kondo model~\cite{Kondo:1964}, but we simply denote this model by ``K."}
The Lagrangian in the $1+1$-dimensional continuous spacetime is given as
\begin{align}
{\cal L} = & {\cal L}_{\chi \mathrm{GN}} + {\cal L}_\mathrm{K}, \label{eq:Lag} \\
{\cal L}_{\chi \mathrm{GN}} = & \bar{\psi} (i \partial\hspace{-0.55em}/-m_l) \psi + \mu \, \bar{\psi} \gamma^{0} \psi  \nonumber \\
& + \frac{G_{ll}}{2N}  \left[ (\bar{\psi} \psi)^2 + (\bar{\psi} i\gamma_5 \psi)^2  \right], \\
{\cal L}_\mathrm{K} = & \bar{\Psi}_v i v^\mu \partial_\mu \Psi_v - \lambda (\bar{\Psi}_v \Psi_v - n_h ) \nonumber \\
& + \frac{G_{hl}}{N} \left[ ( \bar{\psi} \Psi_v ) (\bar{\Psi}_v \psi) + ( \bar{\psi} \gamma^1 \Psi_v )( \bar{\Psi}_v \gamma^1 \psi ) \right],
\end{align}
where $\psi \equiv (\psi_1^T, \ldots, \psi_N^T)$ is a light Dirac fermion field with $N$ components, and the bilinear operators are defined as, e.g., $\bar{\psi} \psi \equiv \sum_{k=1}^N \bar{\psi}_k \psi_k$.
$N \geq 2$ can be regarded as the degeneracy factor from an $SU(N)$-symmetric interaction.
$m_l$ and $\mu$ are the mass and chemical potential of the light fermion, respectively.
$G_{ll}$ is the coupling constant between light fermions, which characterizes condensates composed of only light fermions.\footnote{
The $\chi$GN model at $m_l=0$ satisfies the continuous chiral symmetry, but even at $m_l=0$ the Wilson fermion breaks the chiral symmetry. 
If we are interested in the chiral symmetry in the continuum limit ($a \to 0$) of the W$\chi$GN model, two independent couplings for the scalar and pseudoscalar interactions are required~\cite{Aoki:1985jj}.
The W$\chi$GNK model with two light-light couplings is also straightforward, but in this work we use the same coupling for simplicity.}
$G_{hl}$ is the coupling constant between a light fermion and a heavy fermion, which induces the Kondo effect (or the Kondo condensates).
Note that a non-Abelian interaction  between a light fermion and a heavy fermion is the necessary condition for the Kondo effect.
For example, we can consider a Kondo effect with $N = 2$ mediated by spin, isospin, pseudospin, or $SU(2)$-color exchange and $N = 3$ by $SU(3)$-color exchange as in the usual QCD.
The heavy-fermion field in the HQET is defined as $\Psi_v \equiv \frac{1+ v^\mu \gamma_\mu}{2} e^{im_h vx} \Psi(x)$.
In this form, the original $N$-component Dirac field $\Psi(x)$ at $x^\mu \equiv (t,x^1)$ in real space has a mass $m_h$.
The heavy-fermion velocity $v^\mu$ is set as $v^\mu =(1,0)$, which is the so-called rest frame, and then the phase factor becomes $e^{i m_h t}$.
The original mass term is canceled by this phase factor in the kinetic term, so that it does not appear in the effective Lagrangian.
$\Psi(x)$ is projected into its particle component by the particle projection operator $\frac{1+\gamma_0}{2}$.
$\lambda$ is the Lagrange multiplier for a constraint condition characterizing the heavy-fermion number density with $N$ components defined as $n_h = \bar{\Psi}_v \Psi_v$, and we set $\lambda=0$.

Next, using the Fourier transformation, we get the Lagrangian in momentum space.
In order to get the Lagrangian on the lattice, we replace the spatial momentum $p_1$ in the kinetic term as follows:
\begin{align}
\Slash{p} &= \gamma_0 p_0 - \gamma_1 p_1 \nonumber\\
& \to \gamma_0 p_0 - \frac{1}{a}\gamma_1 \sin{ap_1} - \frac{r}{a} \left( 1-\cos{ap_1} \right), \label{eq:discretization}
\end{align}
where $a$ and $r$ are the lattice spacing and the Wilson parameter, respectively.
The fermion at $r=0$ is called the naive fermion, and $r\neq0$ is the Wilson fermion. 
In what follows, we set $r=1$ and regard that the dimensional quantities are in the lattice unit ($a=1$).

Note that, in our setup, the temporal direction related to $p_0$ is not on the lattice: the space is discretized, but the time is continuous.
This situation corresponds to the usual lattice materials considered in solid-state physics.
On the other hand, for a lattice simulation, the time is also discretized. 
In such a case, one can just replace $p_0$ by the similar form.
Also, in the kinetic term of the heavy fermion, the spatial momentum is zero by taking the rest frame.
Therefore, the heavy-fermion field depends on only $p_0$, so that we need not to replace $p_1$.

Here, we replace the four-point interactions by terms with three types of mean fields: the scalar condensate $\sigma$, pseudoscalar condensate $\Pi$, and Kondo condensates with a gap $\Delta$.
By analogy to the Kondo condensates for the Dirac fermion (the forms without $M$~\cite{Yasui:2016svc,Yasui:2017izi} or with $M$~\cite{Suzuki:2017gde}), we assume the following forms:
\begin{align}
\langle \bar{\psi} \psi \rangle & \equiv - \frac{N}{ G_{ll}} \sigma,   \label{eq:S_con} \\
\langle \bar{\psi} i \gamma_5 \psi \rangle & \equiv - \frac{N}{ G_{ll}} \Pi,  \label{eq:P_con} \\
\langle \bar{\psi} \Psi_v \rangle &\equiv \frac{N}{ G_{hl}} \Delta \sqrt{\frac{E_p + M}{E_p}}, \label{eq:Kondo_con_S} \\ 
\langle \bar{\psi} \gamma^1 \Psi_v \rangle &\equiv  \frac{N}{G_{hl}} \Delta \sqrt{\frac{E_p + M}{E_p}} \frac{-\sin{p_1}+i\Pi}{E_p + M}, \label{eq:Kondo_con_V}
\end{align}
where $E_p$ and $M$ are defined as
\begin{align}
E_p &\equiv  \sqrt{\sin^2 p_1 + M^2 + \Pi^2}, \\
M & \equiv  1 - \cos{p_1} + m_l + \sigma.
\end{align}
The requirement of the two types (scalar and vector types) of Kondo condensates~\cite{Yasui:2016svc,Yasui:2017izi} reflects the particle-component projection for the light Dirac field.
We keep the terms with the condensates, such as $\bar{\psi}\psi \langle \bar{\psi} \psi \rangle$, and neglect the second-order fluctuation terms.
This procedure is equivalent to the large-$N$ limit neglecting fluctuations of auxiliary boson fields.

The resulting mean-field Lagrangian is 
\begin{equation}
{\cal L}_\mathrm{MF} = \bar{\phi} G^{-1}(p_0,p_1) \phi - \frac{N}{2G_{ll}} (\sigma^2 +\Pi^2) - \frac{2N}{G_{hl}} \Delta^2 +\lambda n_h,
\end{equation}
where the inverse propagator of three-component quasiparticle $\phi \equiv (\psi^T, \Psi_v^T)$ in spinor space, composed of the two-component light fermion and the one-component heavy fermion, is
\begin{widetext}
\begin{equation}
G^{-1}(p_0,p_1) = \left(
\begin{array}{ccc}
p_0 + \mu - M             & \sin{p_1}-i\Pi & \Delta^\ast \sqrt{\frac{E_p + M}{E_p}} \\
-\sin{p_1}-i\Pi & -(p_0 + \mu) - M           & -\Delta^\ast \sqrt{\frac{E_p + M}{E_p}} \frac{-\sin{p_1}-i\Pi}{E_p + M} \\
\Delta \sqrt{\frac{E_p + M}{E_p}}  & \Delta  \sqrt{\frac{E_p + M}{E_p}} \frac{-\sin{p_1}+i\Pi}{E_p + M} & p_0 - \lambda \\
\end{array}
\right).
\end{equation}
\end{widetext}
Here, we used the gamma matrices: $\gamma^0=\gamma_0=\sigma_3$, $\gamma^1=-\gamma_1=i\sigma_2$, and $\gamma^5=-\gamma_5=\sigma_1$.
By the diagonalization of the inverse propagator, we obtain the three dispersion relations of the quasiparticles,
\begin{align}
E_\pm(p_1) &\equiv  \frac{1}{2} \left( E_p + \lambda -\mu \pm \sqrt{(E_p-\lambda-\mu)^2 + 8 |\Delta|^2 }\right), \label{eq:Epm} \\
\tilde{E}(p_1) &\equiv - E_p - \mu, \label{eq:Etilde} 
\end{align}
where $E_\pm$ includes the effect of the Kondo condensate $\Delta$, and $\tilde{E}$ is not affected by the Kondo condensate.

After summing up the Matsubara modes from the $p_0$ integral, the thermodynamic potential at inverse temperature $\beta = 1/T$ is written as
\begin{align}
V (\sigma,\Pi,\Delta) =& \nonumber\\
& \hspace{-40pt} \frac{N}{2G_{ll}} (\sigma^2 +\Pi^2) + \frac{2N}{G_{hl}} \Delta^2 - \lambda n_h \nonumber\\
& \hspace{-40pt} - N \int_{-\pi}^\pi \frac{dp_1}{2\pi} \left[\frac{1}{2} \left( \tilde{E} + E_+ + E_- \right) \right. \nonumber\\
& \hspace{-40pt} \left. + \frac{1}{\beta} \ln  \left[ (1+e^{-\beta \tilde{E}}) (1+e^{-\beta E_+}) (1+e^{-\beta E_-}) \right] \right].
\end{align}
By minimizing this potential as a function of $(\sigma,\Pi,\Delta)$, we can estimate the values of $\sigma$, $\Pi$, and $\Delta$.
Note that in this form, since all the terms are proportional to $N$, the values of $\sigma$, $\Pi$, and $\Delta$, as plotted in the next section, do not depend on $N$.
At zero temperature $\beta \to \infty$ and $\lambda=0$, the effective potential is as follows: 
\begin{align}
& V (\sigma,\Pi,\Delta;T\to0,\lambda=0) = \nonumber\\
& \frac{N}{2G_{ll}} (\sigma^2 +\Pi^2) + \frac{2N}{G_{hl}} \Delta^2  - N \int_{-\pi}^\pi \frac{dp_1}{2\pi} \left( -\mu - \tilde{E} - E_- \right).
\end{align}

\section{Numerical results} \label{Sec:3}

\subsection{Massless Wilson fermion}
\begin{figure}[t!]
    \begin{minipage}[t]{1.0\columnwidth}
        \begin{center}
            \includegraphics[clip, width=1.0\columnwidth]{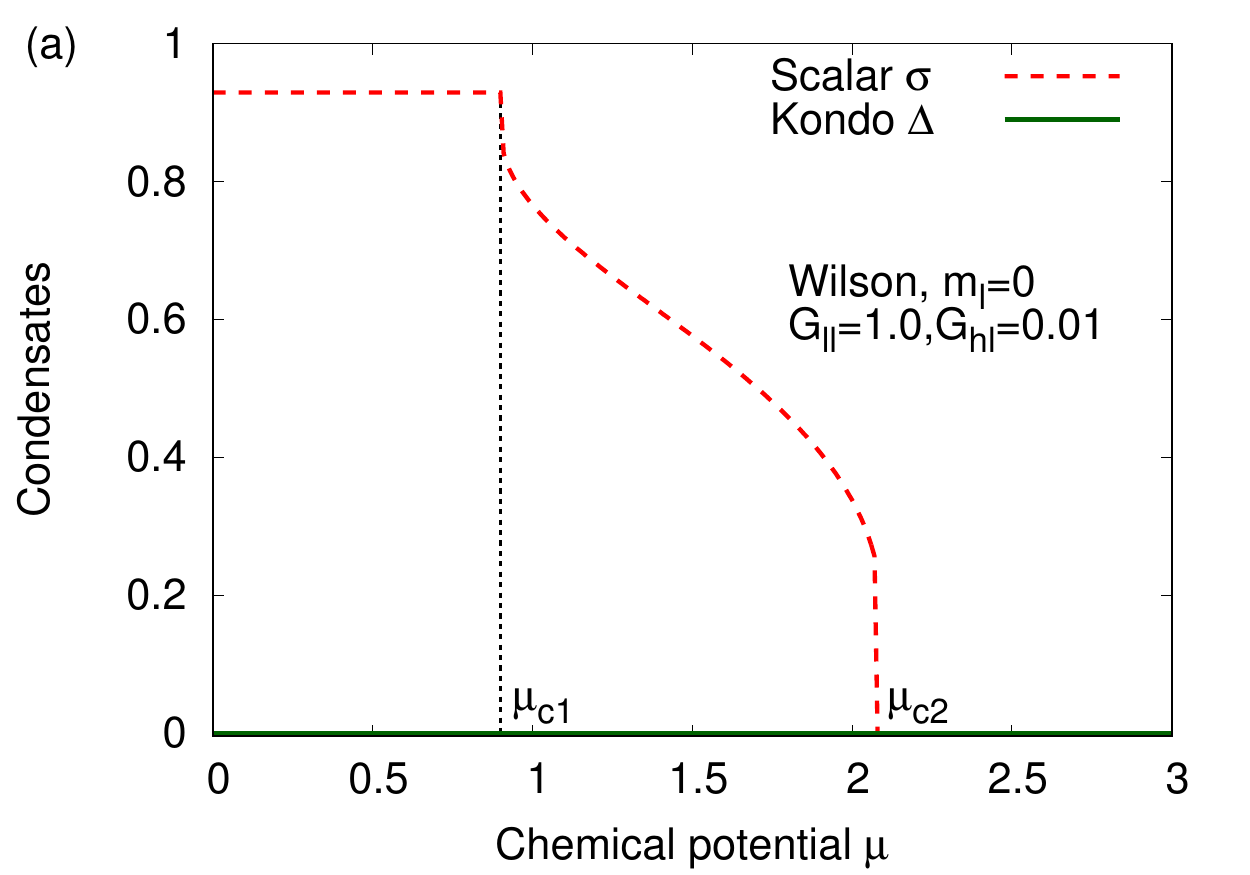}
            \includegraphics[clip, width=1.0\columnwidth]{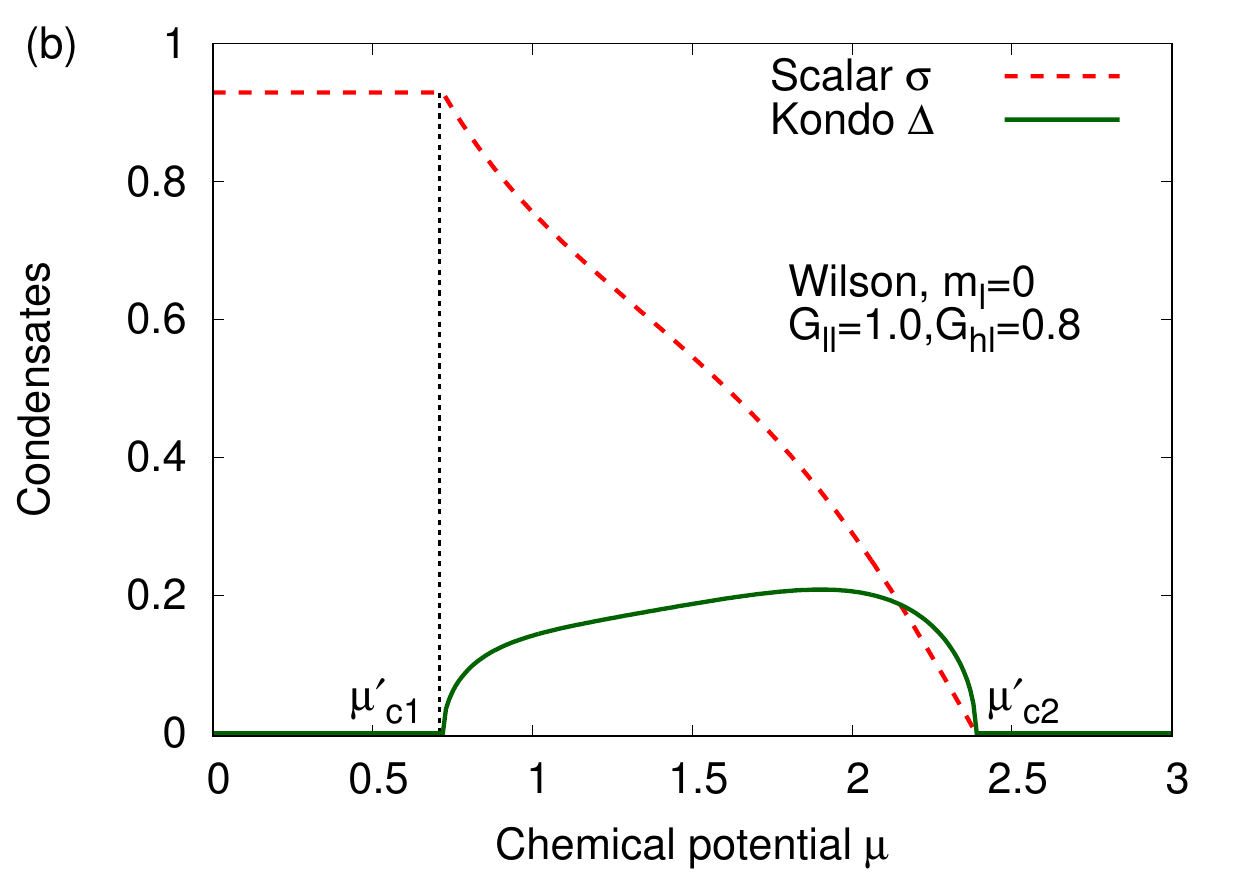}
        \end{center}
    \end{minipage}
    \caption{$\mu$ dependences of $\sigma$ and $\Delta$ for the massless Wilson fermion at (a) a weak heavy-light coupling $G_{hl}=0.01$ and (b) a strong coupling $G_{hl}=0.8$.}
\label{fig:mu_dep}
\end{figure}

First we focus on the Kondo effect for the massless ($m_l=0$) Wilson fermion.
In Fig.~\ref{fig:mu_dep}, we show the $\mu$ dependence of $\sigma$ and $\Delta$ at $G_{ll}=1$.
As shown in Fig.~\ref{fig:mu_dep}(a), when the heavy-light coupling $G_{hl}$ is weak enough, the Kondo effect does not occur, where the phase diagram for $\sigma$ is the same as that in the W$\chi$GN model (without the Kondo effect): we get $\sigma \approx 0.929$ at $\mu=0$. 
From the figure, we find that there are two ``critical" chemical potentials (or transition points), $\mu_{c1} \approx 0.91$ and $\mu_{c2}\approx 2.07$.
$\mu_{c1}$ is the effect from the Fermi level, which is caused by a mechanism similar to the chiral symmetry restoration as in the $\chi$GN model.
$\mu_{c2}$ is the effect from the lattice cutoff (or ultraviolet energy cutoff) for the Wilson fermion, as interpreted in terms of its dispersion relations (see later discussion).

When the heavy-light coupling $G_{hl}$ is strong enough, the Kondo effect occurs, as shown in Fig.~\ref{fig:mu_dep}(b).
In the small-$\mu$ region at $G_{hl}=0.8$, only the scalar condensate $\sigma$ is realized.
In the intermediate-$\mu$ region, the Kondo condensate $\Delta$ appears, and $\sigma$ and $\Delta$ coexist.
Here, as $\mu$ increases, $\sigma$ is gradually reduced, and $\Delta$ increases.
$\mu_{c1}$ for $\sigma$ is shifted to lower $\mu_{c1}^\prime \approx 0.72$ by the appearance of $\Delta$.
Thus, the transitions of both condensates occur at the same time.
Intuitively, some of light fermions in this region start to form the Kondo condensate, and then they do not participate in the formation of the scalar condensate.
As a result, $\mu_{c1}$ is shifted to lower $\mu_{c1}^\prime$ by the appearance of the Kondo condensate.
In the large-$\mu$ region with $\mu \gtrsim 2.39$, we find that both the condensates disappear.
We also point out that $\mu_{c2}$ for $\sigma$ is shifted to higher $\mu_{c2}^\prime \approx 2.39$: the scalar condensate near $\mu_{c2}$ seems to be slightly enhanced by the Kondo effect.
Thus, the shifts of critical chemical potentials, $\mu_{c1}$ and $\mu_{c2}$, would be useful as evidence of the Kondo effect.

\begin{figure}[t!]
    \begin{minipage}[t]{1.0\columnwidth}
        \begin{center}
            \includegraphics[clip, width=0.49\columnwidth]{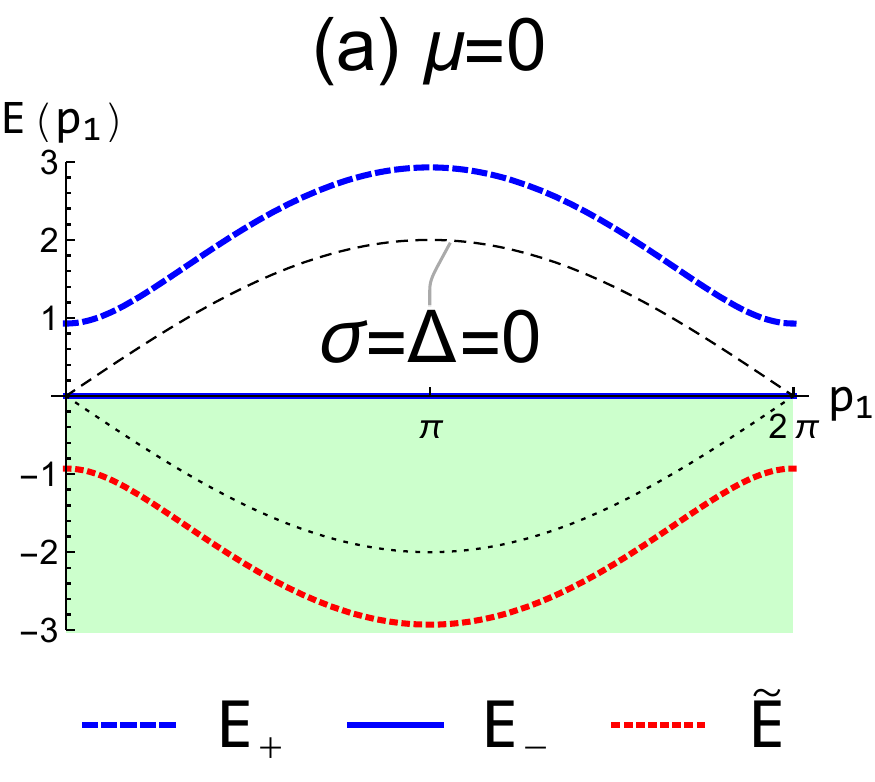}
            \includegraphics[clip, width=0.49\columnwidth]{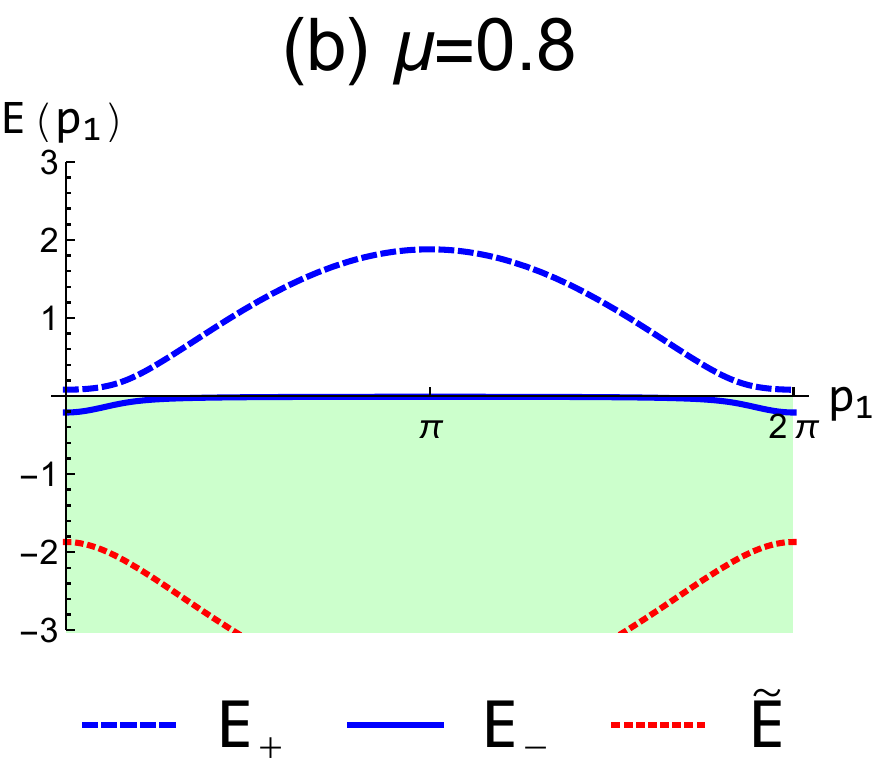}
        \end{center}
    \end{minipage}
    \begin{minipage}[t]{1.0\columnwidth}
        \begin{center}
            \includegraphics[clip, width=0.49\columnwidth]{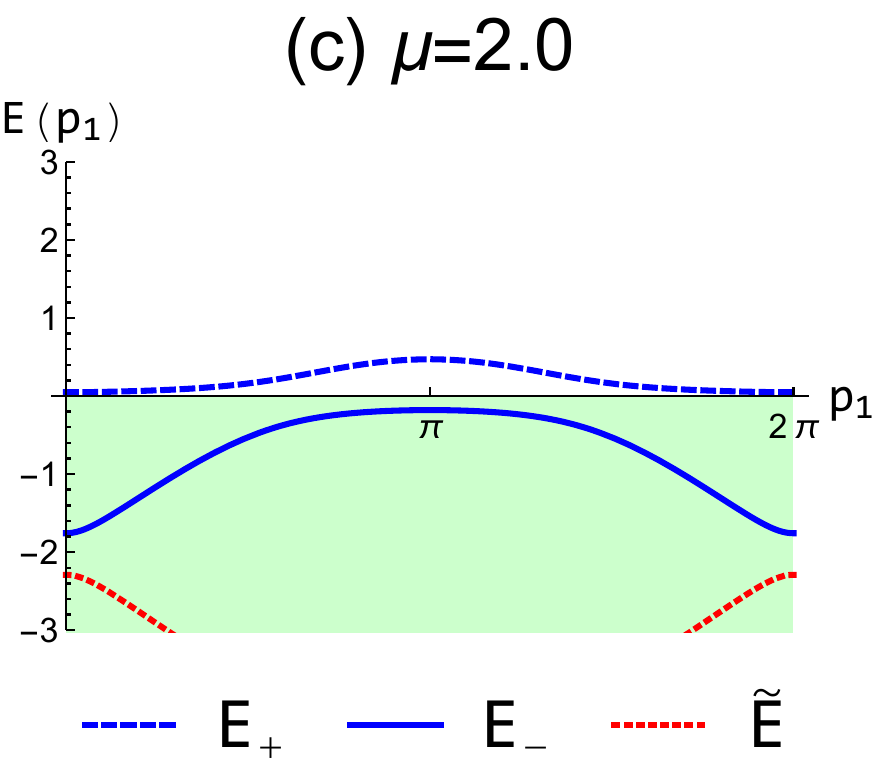}
            \includegraphics[clip, width=0.49\columnwidth]{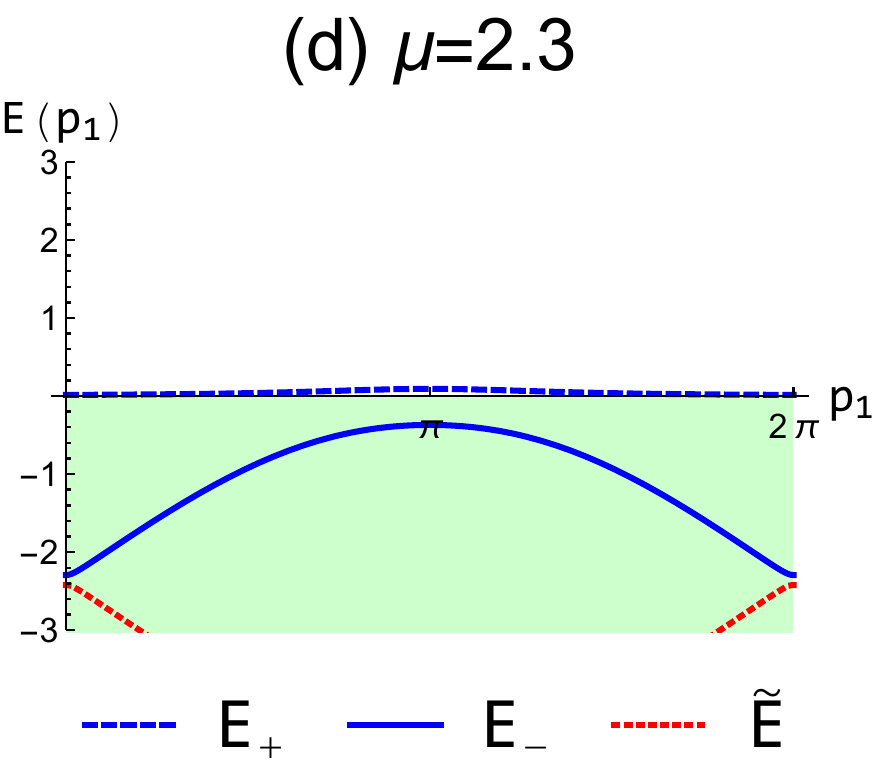}
        \end{center}
    \end{minipage}
    \caption{Dispersion relations of particles at finite $\mu$ and $G_{hl}=0.8$: $E_+$, $E_-$, and $\tilde{E}$.
(a) $\mu=0$, (b) $\mu=0.8$, (c) $\mu=2.0$, (d) $\mu=2.3$.
The black dashed and dotted curves are the Wilson fermions at $\sigma=\Delta=0$.
The colored region means the Dirac or Fermi sea.}
\label{fig:disp}
\end{figure}

In order to interpret our results, in Fig.~\ref{fig:disp}, we show the dispersion relations of the three particles at zero and nonzero $\mu$ at $G_{hl}=0.8$ within the first Brillouin zone, where the explicit forms are given as Eqs.~(\ref{eq:Epm}) and (\ref{eq:Etilde}).
The discussion from the dispersion relations is as follows:
\begin{enumerate}
\item{Small $\mu$}: If $\sigma=\Delta=0$, as plotted as the black dashed and dotted curves in Fig.~\ref{fig:disp}(a), then there is a band crossing point at $p_1=0$, which is the so-called Dirac point in the Wilson fermion.
On the other hand, when $\sigma \neq 0$ in the small-$\mu$ region, the scalar condensate opens a gap between the two dispersions, $E_+$ and $\tilde{E}$, as shown in Fig.~\ref{fig:disp}(a).
The dispersion relation $\tilde{E}$ of the negative-energy band is inside the Dirac sea (equivalently, the Fermi sea at $\mu=0$) and stabilizes the system by the reduction of the free energy.
Note that in this region, $E_-$ is equivalent to the flat band corresponding to the heavy fermion.
Here, the Kondo effect is not realized (unless $G_{hl}$ is large enough).
\item{Intermediate $\mu$}: If $\Delta=0$, with increasing $\mu$, the value of $\sigma$ decreases.
This is because the light-particle dispersion under the Fermi level is occupied, and $\sigma \neq 0$ leads to an enhancement of the free energy, compared to a dispersion with $\sigma=0$.
When the Kondo effect occurs ($\Delta \neq 0$), the light particle and the flat band are mixed by the Kondo condensate.
As a result, $E_-$ inside the Fermi sea stabilizes the system by the reduction of the free energy, as shown in Figs.~\ref{fig:disp}(b) and (c).
\item{Large $\mu$}: In the large-$\mu$ region, the whole dispersion relation of the Wilson fermion is inside the Fermi sea, and the form of the dispersion is not affected by the condensates, as shown in Fig.~\ref{fig:disp}(d).
Note that when $\mu$ is large enough, $E_+$ closely resembles the flat band.
Here, the Kondo effect is not realized, and the heavy fermion on the Fermi level and the massless Wilson fermion inside the Fermi sea are decoupled.
\end{enumerate}

\begin{figure}[tb!]
    \begin{minipage}[t]{1.0\columnwidth}
        \begin{center}
            \includegraphics[clip, width=1.0\columnwidth]{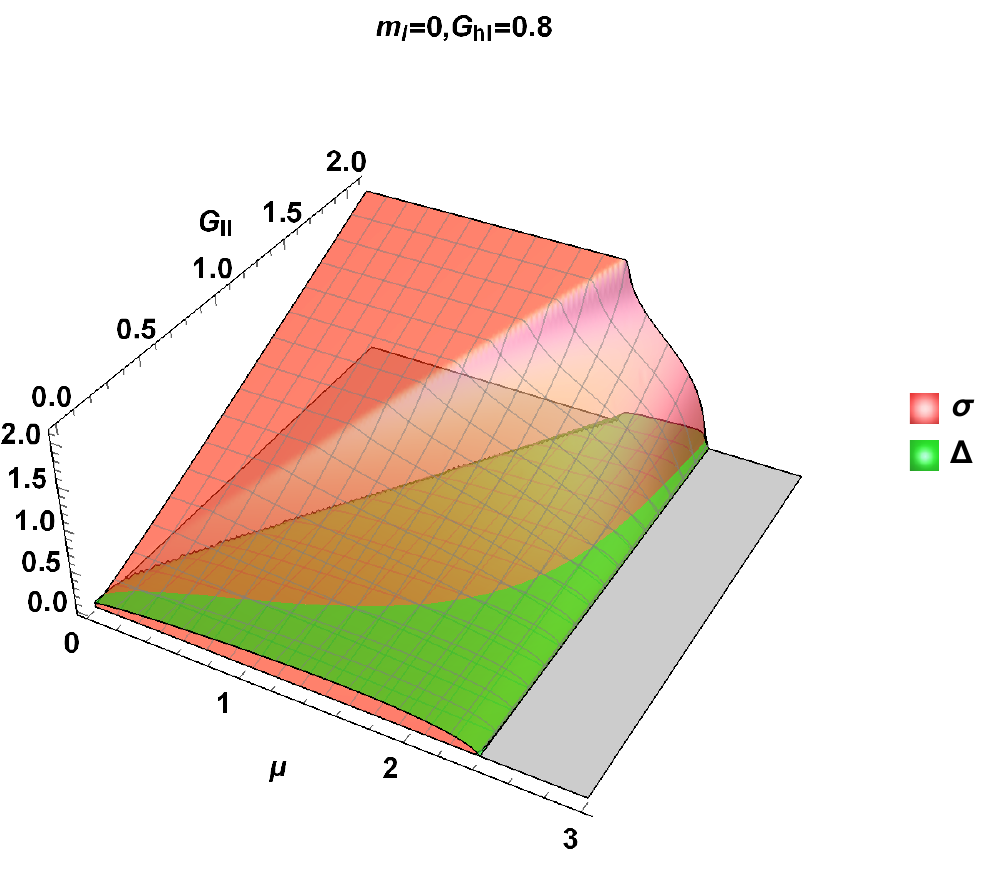}
        \end{center}
    \end{minipage}
    \caption{Phase structure of $\sigma$ and $\Delta$ on the $\mu$-$G_{ll}$ plane at $m_l=0$ and $G_{hl}=0.8$.}
\label{fig:mu_gll}
\end{figure}

Next, we discuss the dependence on the coupling constant between the light fermions, $G_{ll}$. 
In Fig.~\ref{fig:mu_gll}, we show the phase structure on the $\mu$-$G_{ll}$ plane at $G_{hl}=0.8$, where the gray region represents the plane at $\sigma=\Delta=0$.
When $G_{ll}$ is small enough, there is the coexistence phase of the scalar and Kondo condensates.
In the region at large $G_{ll}$ and small $\mu$, the Kondo effect is excluded, and then a pure scalar-condensate phase is realized. 
On the other hand, the region at large $G_{ll}$ and large $\mu$ becomes the coexistence phase (within the plotted region).
Note that in the large-chemical-potential region with $\mu \gtrsim 2.39$, neither of the two condensates can be realized because of the lattice cutoff effect as shown in Fig.~\ref{fig:disp}(d).
Such a second critical chemical potential does not depend on $G_{ll}$.

In \ref{App:A}{Appendix}, we show the results for Kondo effects with the Dirac fermion and naive lattice fermions.

\subsection{Negative-mass Wilson fermion}
Here, we investigate the interplay between the Aoki phase and the Kondo effect.
In the region with a positive mass $m_l > 0$ for the Wilson fermion, the Aoki phase does not appear ($\Pi = 0$).
When a negative mass $m_l < 0$ is switched on, the Aoki phase ($\Pi \neq 0$) can be realized in a parameter region.

\begin{figure}[t!]
    \begin{minipage}[t]{1.0\columnwidth}
        \begin{center}
            \includegraphics[clip, width=1.0\columnwidth]{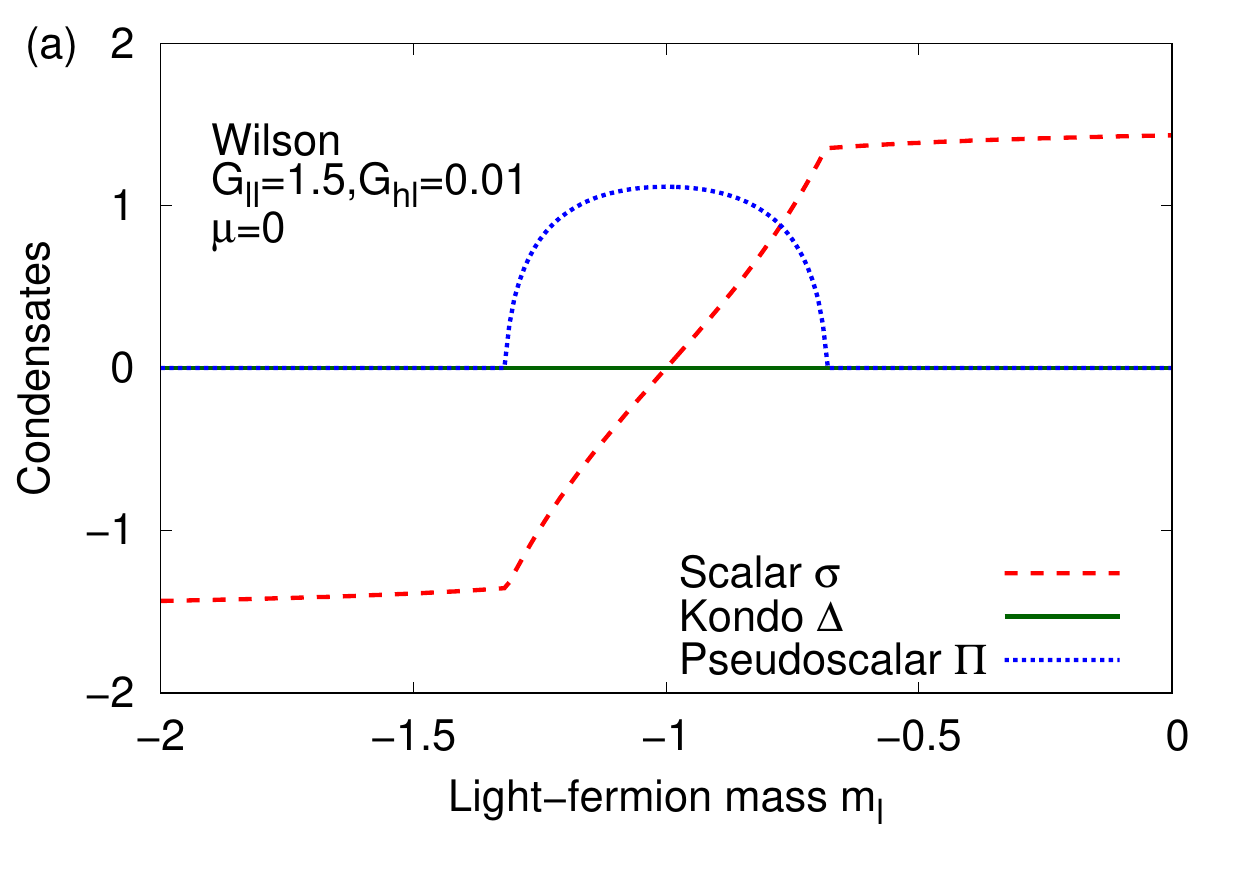}
            \includegraphics[clip, width=1.0\columnwidth]{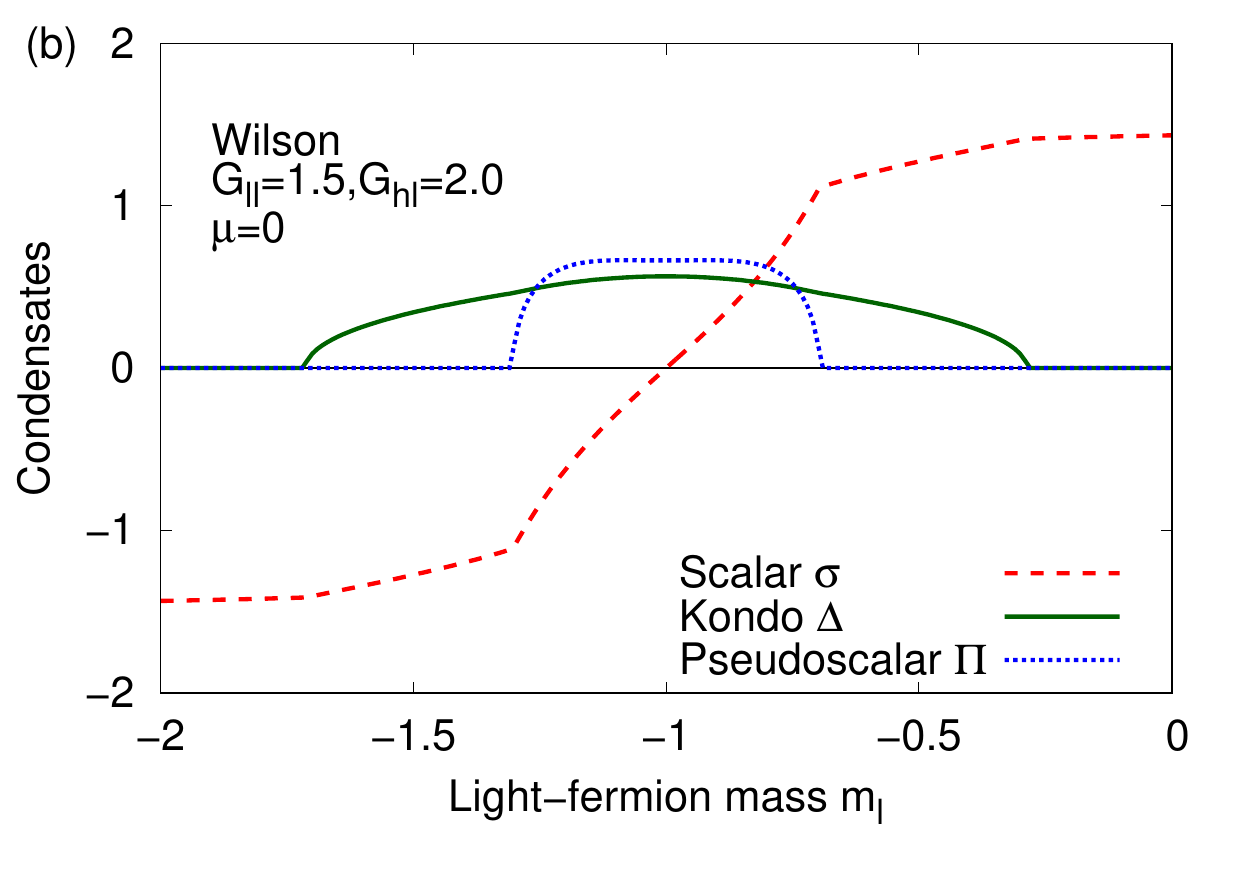}
        \end{center}
    \end{minipage}
    \caption{$m_l$ dependences of condensates for the negative-mass Wilson fermion at (a) a weak heavy-light coupling $G_{hl}=0.01$ and (b) a strong coupling $G_{hl} = 2.0$.}
\label{fig:m0dep}
\end{figure}

In Fig.~\ref{fig:m0dep}, we show the negative-mass dependence of the condensates, where we fixed $G_{ll}=1.5$ to focus on the Aoki phase with a sufficiently large value of the pseudoscalar condensate.
Also, in this figure, we fix $\mu=0$ and change only $G_{hl}$ as a parameter tuning the Kondo effect.
As in Fig.~\ref{fig:m0dep}(a), when the heavy-light coupling is weak enough ($G_{hl}=0.01$), the Kondo effect does not occur, and only the scalar and pseudoscalar condensates are realized.
These behaviors are well known as the conventional Aoki phase scenario in the W$\chi$GN model.

In Fig.~\ref{fig:m0dep}(b) we show the results at $G_{hl}=2.0$.
Here, the heavy-light coupling is large enough, so that the Kondo effect is realized and modifies the other condensates.
From this figure, our findings are as follows:
\begin{enumerate}
\item We find nonzero values of the Kondo condensate around $m_l=-1$, and it coexists with the Aoki phase.
In particular, the Kondo effect is most favored at $m_l=-1$.
This behavior is similar to the Aoki phase in the strong-coupling region.
\item We find that the Kondo effect suppresses both the absolute values of the scalar and pseudoscalar condensates.
Intuitively, this is because light fermions form the Kondo condensate, and then they do not participate in the scalar or pseudoscalar condensate.
Therefore, in experiments or lattice simulations, if one observes such a suppression of the scalar or pseudoscalar condensate, it will be evidence of the Kondo effect.
\end{enumerate}

\begin{figure}[t!]
    \begin{minipage}[t]{1.0\columnwidth}
        \begin{center}
            \includegraphics[clip, width=1.0\columnwidth]{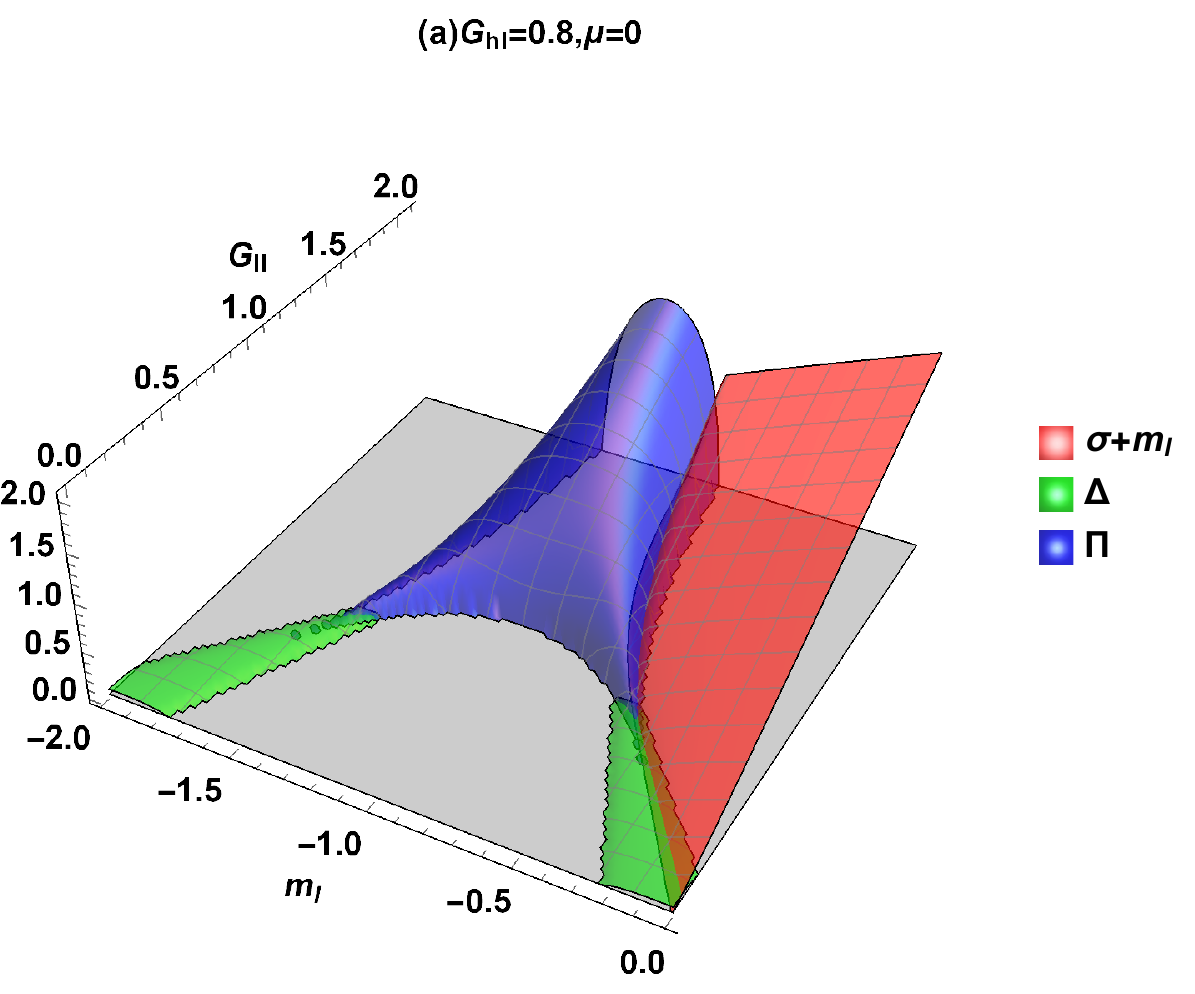}
            \includegraphics[clip, width=1.0\columnwidth]{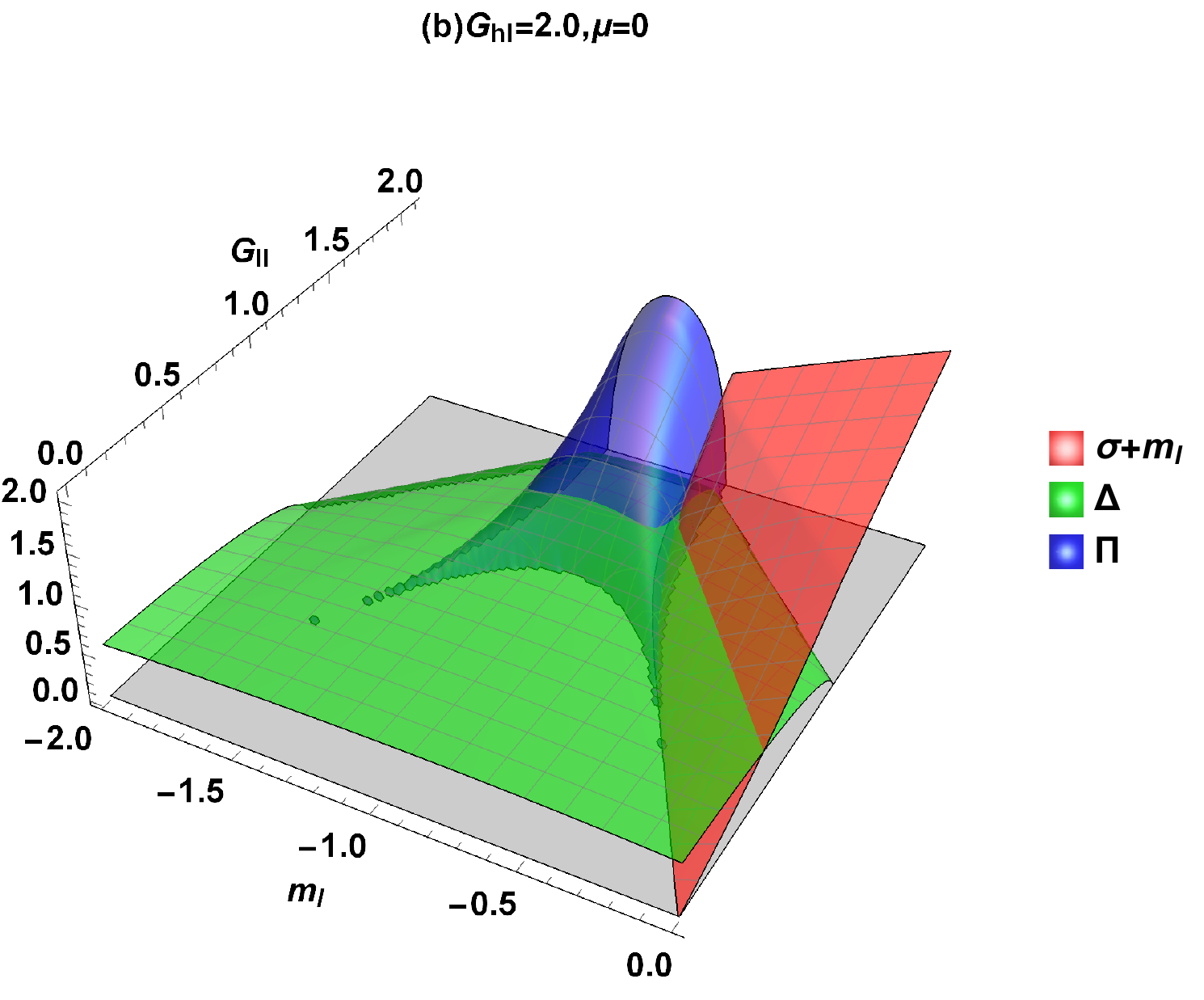}
        \end{center}
    \end{minipage}
    \caption{Phase structure of $\sigma$, $\Delta$, and $\Pi$ on the $m_l$-$G_{ll}$ plane at $\mu=0$ and (a) $G_{hl}=0.8$ or (b) $G_{hl}=2.0$.}
\label{fig:3d_ml-gll}
\end{figure}

In Fig.~\ref{fig:3d_ml-gll}, we show the phase structure of $\sigma$, $\Pi$, and $\Delta$ on the $m_l$-$G_{ll}$ plane at $\mu=0$ and $G_{hl}=0.8$ or $2.0$.
Note that, to improve the visibility, we plot $\sigma + m_l$ instead of $\sigma$.
From this figure, we find that $\Delta$ is favored in the weak $G_{ll}$ region.
In particular, at $G_{hl}=0.8$, the Kondo condensate appears only in the parameter regions that should have been the Aoki phase at $G_{hl}=0$ (the so-called Aoki fingers or cusp region).
This is because the dispersion relation at $m_l=0$ or $m_l=-2$ is gapless at $p_1=0$ or $p_1=\pi$, and such a gapless band is closer to the flat band of the heavy fermion.
On the other hand, the dispersion relation at $0<m_l<-2$ is gapped at any momentum, which is away from the heavy-fermion band.
In the region with the Kondo condensate, the value of the pseudoscalar condensate is suppressed, and the original Aoki phase can be excluded by the Kondo effect. 
In other words, the Aoki fingers are covered by ``fingernails" of the Kondo condensate phase.
Furthermore, at stronger heavy-light coupling ($G_{hl}=2.0$), we find a wide region of the Kondo condensate, which expands to larger-$G_{ll}$ region.
Thus, our results indicate that the effects from heavy impurities (via $G_{hl}$) can be significant in weakly coupling region for $G_{ll}$. 

\begin{figure}[tb!]
    \begin{minipage}[t]{1.0\columnwidth}
        \begin{center}
            \includegraphics[clip, width=1.0\columnwidth]{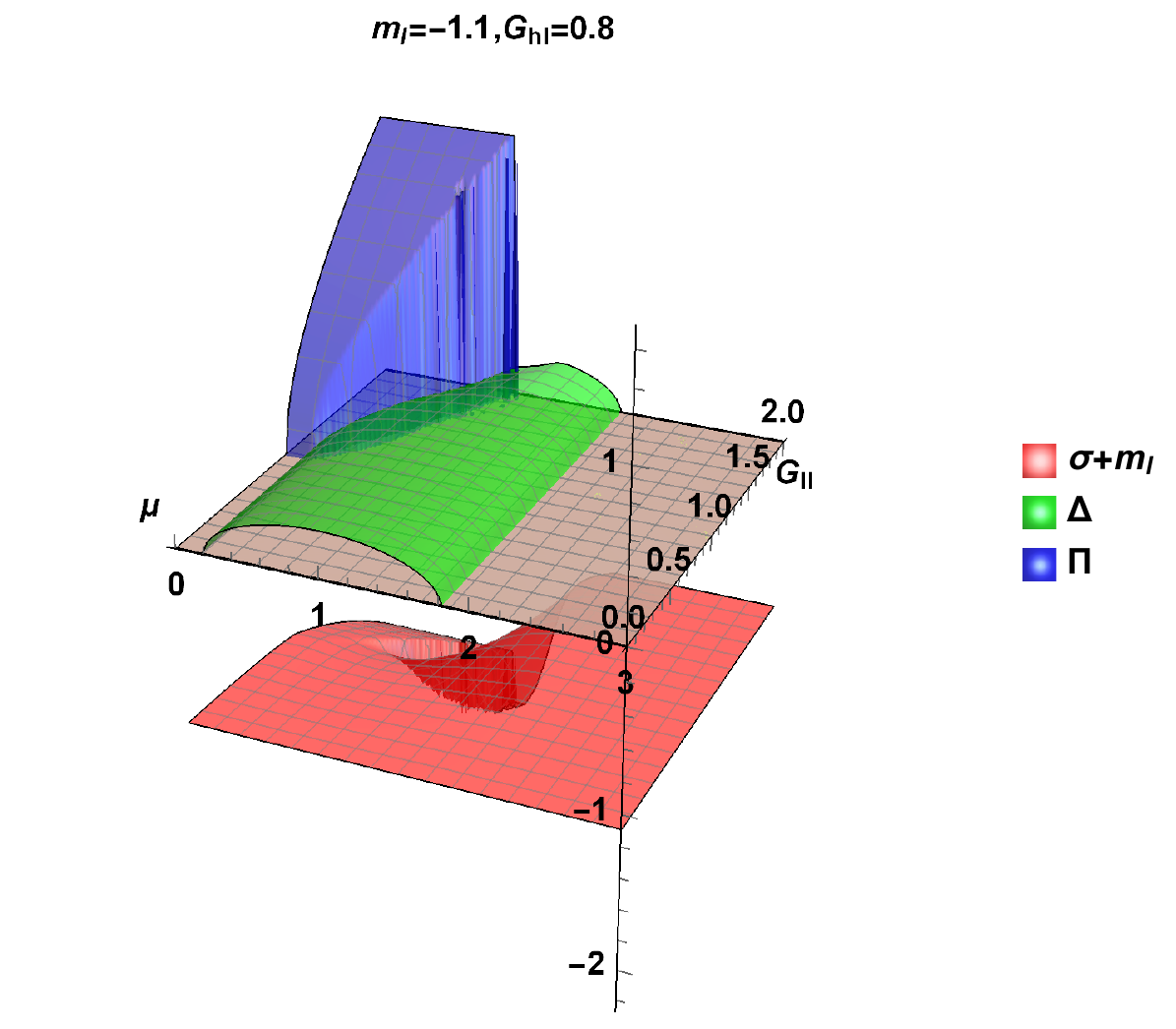}
        \end{center}
    \end{minipage}
    \caption{Phase structure of $\sigma$, $\Delta$, and $\Pi$ on the $\mu$-$G_{ll}$ plane at $m_{l}=-1.1$ and $G_{hl}=0.8$.}
\label{fig:mu_gll_ml11}
\end{figure}

In order to discuss the $\mu$ dependence, in Fig.~\ref{fig:mu_gll_ml11}, we show the phase structure on the $\mu$-$G_{ll}$ plane at $G_{hl}=0.8$ and $m_{l}=-1.1$.\footnote{Note that $m_l=-1$ is a special parameter because the dispersion relations become two flat bands (for light fermions).
Such a situation is interesting as physics of the flat band (e.g., see Refs.~\cite{Junemann:2016fxu,Ishikawa:2020icy}), but we focus on the Wilson fermion at $m_{l}=-1.1$ in the main text.}
As $\mu$ increases, the pseudoscalar condensate decreases, while the Kondo condensate increases.
Around the transition point of $\mu$, the scalar condensate is also modified.
Similar to Fig.~\ref{fig:mu_gll} at $m_l=0$, when the chemical potential is large enough, all condensates are zero by the lattice cutoff effect.

We emphasize that the negative-mass region of the Wilson fermion can be regarded as an effective model for the bulk of topological insulators.\footnote{In particular, the weak-coupling region in 1+1 dimensions for odd $N$ is a topological insulator belonging to the symmetry class BDI~\cite{Bermudez:2018eyh}, where a topological invariant is characterized by the Zak phase defined as the integral of the Berry connection.}
The two parameters in the Hamiltonian of the Wilson fermion, the mass $m_l < 0$ and the Wilson parameter $r$, can be related to the band structure of a material, which is determined by the original band and the strength of the spin-orbit interaction.
An intrinsic spin-orbit interaction may be roughly tuned by changing the chemical composition of the material (e.g., for BiTl(S${}_{1-\delta}$Se${}_{\delta})_2$, see Ref.~\cite{Xu:2011dc}).
In this sense, one can experimentally examine the negative-mass dependence.
As in Fig.~\ref{fig:3d_ml-gll}(a), we have found that the phase transitions (namely, appearance or disappearance) of the Kondo condensate $\Delta$ significantly depend on the negative mass $m_l$.
Therefore, experimentally, one could capture such a phase transition by tuning the spin-orbit interaction.

On the other hand, for topological insulators, the coupling constant (corresponding to $G_{ll}$) between electrons is usually small, so that the parameter region with the Aoki phase may be narrow.
Instead of topological insulators in the strong-coupling region, the axion insulators (see, e.g., Ref.~\cite{Sekine:2021}) are other candidates to study the interplay between a parity-symmetry breaking ground state and the Kondo effect.
In order to build an effective model to describe axion insulators, we have to introduce a pseudoscalar-mass term such as $i m_5 \bar{\psi} \gamma_5 \psi$.
Investigation of Kondo effects based on such an effective model will be straightforward.
From our results shown in this paper, we can expect impurity effects in axion insulators by regarding the pseudoscalar condensate $\Pi$ as the pseudoscalar mass $m_5$.
For example, we can expect the appearance of the Kondo effect at a small $m_5$ and the suppression of the Kondo effect by a large $m_5$.

\section{Conclusion and outlook} \label{Sec:4}
In this paper, we have investigated the Kondo effect for the Wilson fermion with the four-point interaction, which is based on the discretization (\ref{eq:discretization}) of the $\chi$GNK model (\ref{eq:Lag}).
From our model, we have found (i) a coexistence phase of the Kondo condensate and other condensates such as the scalar and pseudoscalar condensates, (ii) a shift of the critical chemical potential of the scalar condensate by the Kondo effect, and (iii) an interplay between the Kondo effect and Aoki phase (particularly, the Kondo fingernails structure).

It should be noted that our W$\chi$GNK model is a choice of models describing the Kondo effect for the Wilson fermion, and other types of W$\chi$GNK models may be also constructed.
For example, we have used the heavy-fermion field based on the {\it leading order} of HQET, but the building of W$\chi$GNK models based on its higher orders or heavy Dirac fermions will be also interesting.
Furthermore, the mean-field assumptions might be improved.
We have assumed the condensates (\ref{eq:S_con})--(\ref{eq:Kondo_con_V}), but other types of light-fermion condensates and Kondo condensates, e.g., including spatially inhomogeneous condensates, might be possible.
Such a detailed examination is left for future studies.
Also, it will be interesting to extend our model to higher spatial dimensions, such as the NJL$_3$ and NJL$_4$ models, or to replace the four-point interactions by other interactions, such as non-Abelian gauge interactions.

In this work, we have focused only on the situations with a single light flavor (the number of flavors is not $N$ but $N_f$), which will be examined by $N_f=1$ lattice simulations.
We comment on the extension to the $N_f=2$ case.
In this case, additional flavor degrees of freedom may lead to ``overscreening" of the Kondo effect, and non-Fermi-liquid behavior can appear, which is the so-called multichannel Kondo effect~\cite{Nozieres:1980} (see Refs.~\cite{Kanazawa:2016ihl,Kimura:2016zyv} for expectations for the QCD Kondo effect).
In such a situation, the standard mean-field approximation may be useless, and then one has to use an alternative approach, such as $N_f=2$ lattice simulations.
As a direct measurement for the Kondo effect in lattice simulations, one may measure the vacuum expectation value of a heavy-light bilinear operator, such as $\langle \bar{\psi} \Psi_v \rangle$ defined in this paper.
Also, the values of light-fermion condensates $\langle \bar{\psi} \psi \rangle$ and $\langle \bar{\psi} i\gamma_5 \psi \rangle$ are modified by the Kondo effect, and they will be indirect evidence of the Kondo effect.
Furthermore, heavy-light mesonic two-point correlators also could be influenced by the Kondo effect.

Monte Carlo simulations of the W$\chi$GN model at finite chemical potential may suffer from the sign problem.
In this case, one can expect the realization of the Kondo effect by tuning the heavy-light coupling constant.
Also, even at finite chemical potential, sign-problem-free approaches, such as the tensor renormalization group~\cite{Takeda:2014vwa}, the matrix product state~\cite{Bermudez:2018eyh}, and the projected-entangled-pair state~\cite{Ziegler:2020zkq}, will be useful.

In addition, cold-atom simulations may be also promising candidates for examining both the interacting Wilson fermion \cite{Bermudez:2010da,Mazza:2011kf,Kuno:2018rmc,Zache:2018jbt} and the Kondo effect, e.g.,~\cite{Paredes:2003,Duan:2004,Gorshkov:2010,Foss-Feig:2010,Foss-Feig:2010b,Bauer:2013,Nishida:2013,Nakagawa:2016,Nakagawa:2018}, where tuning the coupling constants rather than the chemical potential will be useful for elucidating the Kondo effect.

\section*{ACKNOWLEDGMENTS}
The authors thank Yasufumi Araki, Daiki Suenaga, and Shigehiro Yasui for helpful discussions.
This work was supported by Japan Society for the Promotion of Science (JSPS) KAKENHI (Grants No. JP17K14277 and No. JP20K14476).
T.I. was supported by RIKEN Junior Research Associate Program.

\appendix*
\section{Kondo effects for Dirac and naive fermions} \label{App:A}

\begin{figure}[t!]
    \begin{minipage}[t]{1.0\columnwidth}
        \begin{center}
            \includegraphics[clip, width=1.0\columnwidth]{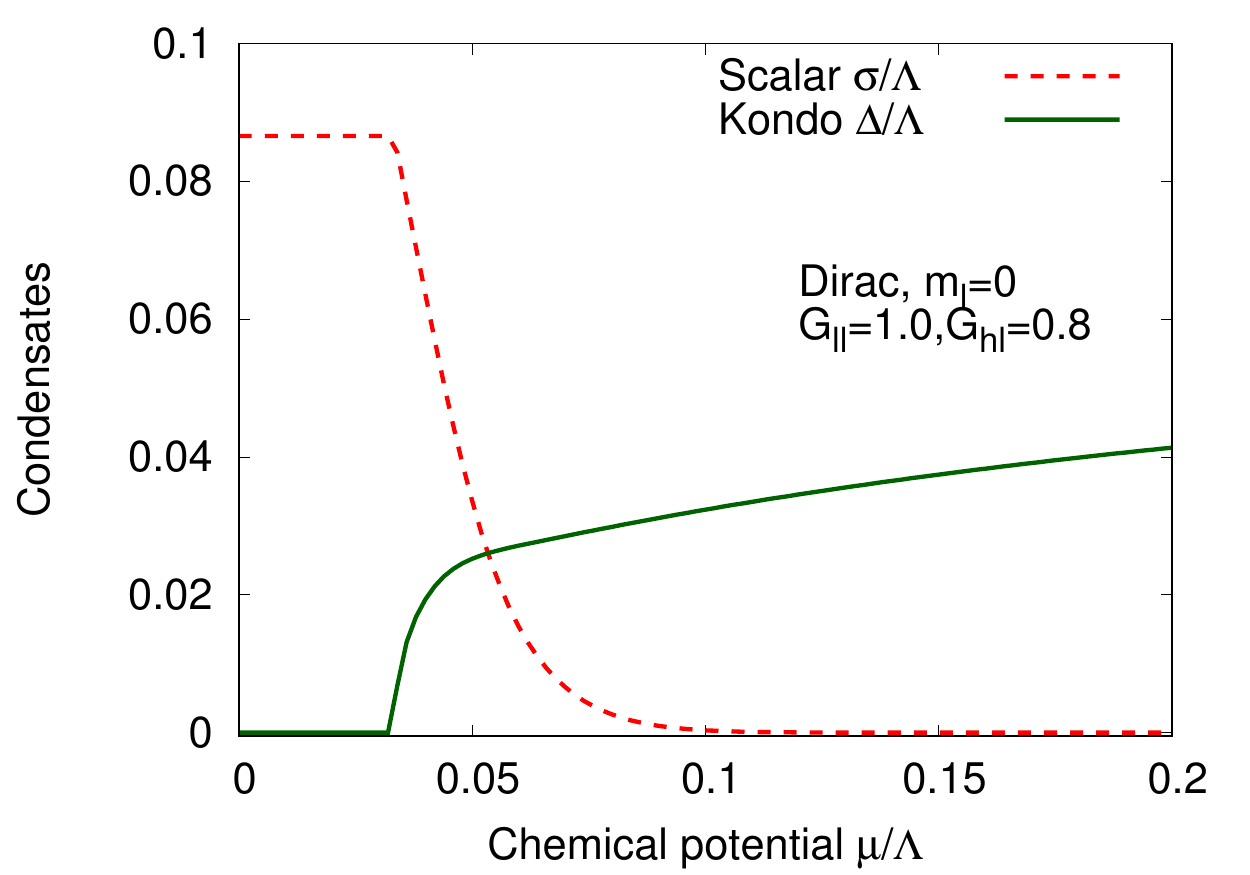}
        \end{center}
    \end{minipage}
    \caption{$\mu$ dependences of $\sigma$ and $\Delta$ for the massless Dirac fermion at a strong heavy-light coupling $G_{hl}=0.8$.}
\label{fig:mu_Dirac}
\end{figure}

\begin{figure}[t!]
    \begin{minipage}[t]{1.0\columnwidth}
        \begin{center}
            \includegraphics[clip, width=1.0\columnwidth]{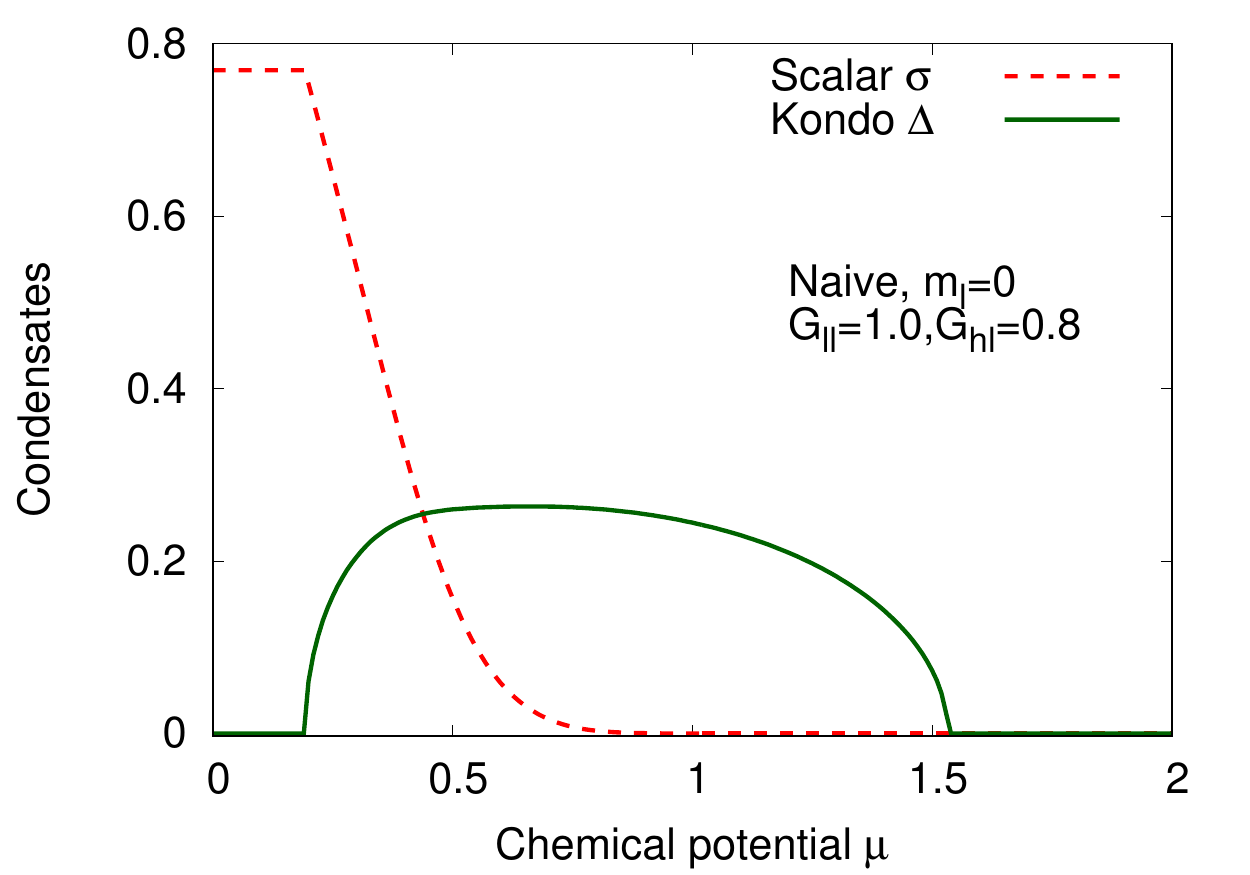}
        \end{center}
    \end{minipage}
    \caption{$\mu$ dependences of $\sigma$ and $\Delta$ for the massless naive fermions at a strong heavy-light coupling $G_{hl}=0.8$.}
\label{fig:mu_naive}
\end{figure}

In this Appendix, we qualitatively compare Kondo effects for other fermions with that for the Wilson fermion.
Here we focus on the Dirac fermion and naive lattice fermion in $1+1$ dimensions: we investigate the phase structures of the ``Dirac-chiral-Gross-Neveu-Kondo model," defined as the Lagrangian~(\ref{eq:Lag}) and the ``naive-chiral-Gross-Neveu-Kondo model," defined using the discretization (\ref{eq:discretization}) at $r=0$.

In Fig.~\ref{fig:mu_Dirac}, we show the results for the Dirac fermion, where the momentum integral interval in the effective potential is $- \Lambda \leq p_1 \leq \Lambda$ with a cutoff $\Lambda$.\footnote{In $1+1$ dimensions, where the coupling constants are dimensionless, the $\mu/\Lambda$ dependences of dimensionless condensates regularized by the cutoff $\Lambda$ do not depend on the value of the cutoff.}
In the intermediate-$\mu$ region, the coexistence phase of $\sigma$ and $\Delta$ appears, which is similar to that of the Wilson fermion.
In the large-$\mu$ region, $\sigma$ becomes zero whereas $\Delta$ survives.
Such behavior is distinct from the case of the Wilson fermion in which $\sigma$ and $\Delta$ becomes zero at the same time because of the lattice cutoff.

In Fig.~\ref{fig:mu_naive}, we also show the results for the naive fermion.
The behavior in the intermediate-$\mu$ region is similar to the Wilson and Dirac fermions.
In the large-$\mu$ region, $\sigma$ becomes zero.
At higher $\mu$, $\Delta$ also becomes zero by the lattice cutoff effect.
Thus, the critical chemical potentials for $\sigma$ and $\Delta$, $\mu_{c\sigma}$ and $\mu_{c\Delta}$, are different.
This is different from the Wilson fermion, where $\mu_{c\sigma}$ and $\mu_{c\Delta}$ are almost the same.

Note that, for both the Dirac and naive fermions, the phase transition of $\sigma$ at finite $\mu$ without the Kondo effect is first order, where the order parameter $\sigma$ discontinuously drops to zero.
On the other hand, when the Kondo effect is switched on, its order is smeared.

\bibliography{reference}
\end{document}